%% file: sample-sigconf.tex
\documentclass[sigconf]{acmart}

\usepackage{hyperref}
\usepackage[font=small,labelfont=bf,tableposition=top]{caption}
\usepackage{graphicx}
\usepackage{amsmath, amssymb}
\usepackage[ruled]{algorithm2e}
\usepackage{subfigure}
\usepackage{multirow}
\usepackage{caption}
\usepackage{accents}
\usepackage[subtle,tracking=normal,margins=normal,leading=normal,title=tight]{savetrees}

\DeclareCaptionLabelFormat{andtable}{#1~#2  \&  \tablename~\thetable}

\AtBeginDocument{%
  \providecommand\BibTeX{{%
    \normalfont B\kern-0.5em{\scshape i\kern-0.25em b}\kern-0.8em\TeX}}}

\acmYear{2020}\copyrightyear{2020}
\setcopyright{acmcopyright}
\acmConference[PODC '20]{Symposium on Principles of Distributed Computing}{August 3--7, 2020}{Virtual Event, Italy}
\acmBooktitle{Symposium on Principles of Distributed Computing (PODC '20), August 3--7, 2020, Virtual Event, Italy}
\acmPrice{15.00}
\acmDOI{10.1145/3382734.3405704}
\acmISBN{978-1-4503-7582-5/20/08}

\newcommand{\pp}{p2p } 
\newcommand{\name}{Perigee } 
\newcommand{\namens}{Perigee} 
\newcommand{\ie}{{\em i.e., }}
\newcommand{\dbtilde}[1]{\accentset{\approx}{#1}}

\newtheorem{thm}{Theorem}

\begin{document}

\title{\namens: Efficient Peer-to-Peer Network Design for Blockchains}

%
\author{Yifan Mao}
\affiliation{%
  \institution{The Ohio State University}}
\email{mao.360@osu.edu}

\author{Soubhik Deb}
\affiliation{%
  \institution{University of Washington Seattle}}
\email{soubhik@uw.edu}

\author{Shaileshh Bojja Venkatakrishnan}
\affiliation{%
  \institution{The Ohio State University}}
\email{shaileshh.bv@gmail.com}

\author{Sreeram Kannan}
\affiliation{%
  \institution{University of Washington Seattle}}
\email{ksreeram@uw.edu}

\author{Kannan Srinivasan}
\affiliation{%
  \institution{The Ohio State University}}
\email{kannan.sriniv@gmail.com}
%
%
%
%
%
%

\renewcommand{\shortauthors}{Mao, et al.}

\begin{abstract}
A key performance metric in blockchains is the  latency between when a transaction is broadcast and when it is confirmed (the so-called, confirmation latency).  While improvements in consensus techniques can lead to lower confirmation latency, a fundamental lower bound on confirmation latency is the propagation latency of messages through the underlying peer-to-peer (p2p) network (in Bitcoin, the propagation latency is several tens of seconds).  The {\em de facto} p2p protocol used by Bitcoin and other blockchains is based on random connectivity: each node  connects to a random subset of nodes. The induced p2p network topology can be highly suboptimal since it neglects geographical distance, differences in bandwidth, hash-power and computational abilities across peers. 
We present \namens, a decentralized algorithm that automatically learns an efficient p2p topology tuned to the aforementioned network heterogeneities, purely based on peers' interactions with their neighbors.  
Motivated by the literature on the multi-armed bandit problem, \name optimally balances the tradeoff between retaining connections to known well-connected neighbors, and exploring new connections to previously-unseen neighbors. 
Experimental evaluations show that \name reduces the latency to broadcast by $33\%$.  
Lastly \name is simple, computationally lightweight, adversary-resistant, and compatible with the selfish interests of peers, making it an attractive p2p protocol for blockchains. 
\end{abstract}




\maketitle

\input{intro}

\input{model}

\input{baseline}

\input{design}

\input{eval}

\input{conclusion}

%
%
%


\bibliographystyle{ACM-Reference-Format}
\bibliography{sample-base}


\end{document}

%% file: intro.tex
\section{Introduction} \label{s:intro}
In $2008$, Satoshi Nakamoto proposed Bitcoin as a decentralized currency system over a peer-to-peer (p2p) network, with the blockchain protocol as its underlying technology for maintaining a public ledger of payment transactions~\cite{nakamoto2019bitcoin}.
Since then, there has been a proliferation of applications leveraging the power of blockchains as a core component, for implementing cryptocurrencies, smart contracts,
supply chain management, etc. ~\cite{cong2019blockchain}. 
Today, the combined market capitalization of all cryptocurrencies is around 280 billion dollars with a rapidly increasing user base~\cite{bitcoinusage}.  

A key problem facing blockchain systems today is scalability---for example, the Bitcoin network can currently support a maximum of only 10 transaction per second~\cite{sompolinsky2015secure}, compared to 1700 transactions per second on Visa. 
A blockchain protocol functions by periodically consolidating transactions and broadcasting them as ``blocks'' over the network. 
Recent works have constructed new consensus protocols to improve confirmation latency and throughput in both the permissioned ~\cite{pass2017hybrid,abraham2019sync} as well as the permissionless settings ~\cite{bagaria2019prism, Gilad2017algorand}. 
There have also been methods to 
compress ~\cite{ozisik2017graphene} and code \cite{chawla2019velocity} blocks while forwarding. 
Despite these improvements, a  fundamental factor limiting the performance of blockchain systems is the inherent message propagation delay introduced by the p2p network. 
A block experiences delays from various factors during propagation, such as due to link latencies and processing delays for verifying blocks at each peer. 
It is known that improving the propagation delay directly improves key performance metrics of the system: transaction throughput, latency in confirming transactions, and security~\cite{bagaria2019prism}. 

Measurement studies over Bitcoin~\cite{croman2016scaling, decker2013information} report that it takes on an  average 79 seconds for a block to reach 90\% of nodes in the network. 
Whereas the median round-trip-time between hosts on the Internet is <300ms~\cite{hoiland2016measuring}, the median bandwidth of Bitcoin nodes is 33Mbps ~\cite{croman2016scaling}, and the average time taken to validate a block is  <200ms~\cite{gervais2015tampering}. 
Blocks have an average size of 1MB in Bitcoin today.
These numbers show that the time it takes for a block to propagate to the majority of the network is $40\times$ larger than the time it takes to verify and relay a block between two nodes (<1 or 2s). 
With an estimated number of less than 11,000 nodes in Bitcoin~\cite{bitnodescom}, and each node making connections to at least 8 other nodes~\cite{miller2015discovering}, a key reason for the disproportionately large propagation delay today is due to the ineffective way in which nodes are connected to each other (\ie the topology) in the p2p network.  

The design of p2p networks for efficient content storage and lookup has a long history \cite{lua2005survey, stoica2001chord, rowstron2001pastry}. However, blockchains require only a simple broadcast primitive (for example, unicast messages directed to a particular node or lookup for specific content are not supported) and this primitive needs to be robust to adversarial action. This has led to Bitcoin following a random connection policy, where each node chooses its neighbors randomly from among a set of known nodes. 
While the random graph topology is simple, robust and provides good connectivity (from a graph theoretic standpoint), it is oblivious to differences in round-trip-time latencies between different nodes, heterogeneity in node bandwidth and block verification times. 
This inherently worsens the overall delay experienced by broadcasted blocks; for example, a block is likely to make several back-and-forth trips across distant continents before reaching a node.
In this paper, we consider the question of how to optimally connect nodes in the Bitcoin network (and blockchain p2p networks in general), in a way that is aware of link and node heterogeneities, so that the broadcast time of blocks is minimized. 

We present \namens, a decentralized protocol that adaptively decides which neighbors a node should connect to, purely based on the node's past interactions with its neighbors. 
Our protocol is motivated by the classical multi-armed bandit problem~\cite{auer2002using}.
Nodes in \name balance the trade-off between retaining old neighbors with good connectivity, and exploring new neighbors with potentially better connectivity.
In \namens, a node quantifies its interactions with its neighbors by looking at the block arrival times. 
Neighbors that consistently deliver blocks quickly are favored, while others are disconnected.  
\name also continuously forms connections to a small number of nodes randomly, for discovering previously unseen but well-connected nodes. 
Our approach of purely using block arrival times to select neighbors is automatically tuned to heterogeneity in link latencies, block validation delays and node bandwidth. 
The end result is a topology that is very tight: experimental results show \name improves overall propagation delay by 33\% compared to the state of the art (\S\ref{s:eval}).

Modifying the p2p topology for faster block propagation has been considered in prior works; e.g., in Kadcast~\cite{rohrer2019kadcast} the authors propose a structured p2p overlay 
as a faster alternative to the random topology.  
However, such a structured topology is still oblivious to link latencies, block validation times and node bandwidth, which renders its performance to be only slightly better than the random topology (\S\ref{s:eval}). 
One way to take link latency into account is by using the geographical location of nodes, inferred from their IP addresses, to select neighbors
~\cite{abboud2009underlay}. 
However, this approach does not accurately reflect propagation latencies since nodes frequently use proxy-servers, VPN and Tor to run nodes, not to mention potential geo-location spoofing attacks by adversaries. 
Even more importantly, this approach also remains oblivious to the differing processing power and 
bandwidth possessed by different nodes. 

In contrast, \name does not use any explicit property about a node, and is thus much more robust to spoofing attacks. 
Another line of work proposes high-speed block distribution networks (e.g., BloXroute~\cite{klarman2018bloxroute}, Falcon~\cite{falcon}, Fibre~\cite{fibre}) 
to reduce block propagation times. 
These solutions are not fully decentralized, as using them requires nodes to place trust on the relay network. 
The routes in these networks could be also susceptible to man-in-the-middle attack. 
Nevertheless, even if such relay networks are present, \name automatically adapts its topology to exploit those networks (\S\ref{s:fastdistnet}). 




\name naturally incentivizes nodes to follow protocol---if a node deviates from protocol (e.g., stops relaying blocks, or does not update its neighbors using \namens), then its neighbors will penalize the node by disconnecting from it in the future. Consequently, the deviant node will lose out on receiving blocks in a timely manner.  

Finally \name maintains a subset of random neighbors at all times, thus making it less susceptible to eclipse attacks.


\subsection{Background} \label{s: background}

Blockchain applications use a distributed, replicated ledger---called the blockchain---for storing and updating, collective states of application's end-users. 
Bitcoin is a popular example of a blockchain application. 
In Bitcoin, the blockchain contains the sequence of all payment transactions made by users since the very beginning of Bitcoin. 
The public nature of these transaction logs allows a payee to unilaterally verify the authenticity of incoming payments, without relying on third party organizations. 
Thus it is a fully decentralized payment system, a property that has contributed significantly to its growth and popularity.

\subsubsection{Bitcoin architecture} 
Bitcoin operates over a p2p network. 
In Bitcoin, when a user makes a payment transaction, first a transaction message specifying the sender, recipient, and payment amount is created by the user.  
The transaction is then broadcast to other peers over the network. 
As new transactions are propagated over the network, special peers called miners accumulate these transactions, verify their authenticity and consolidate them into individual transaction {\em blocks} in a process called mining. 
A block in Bitcoin can contain a few thousand transactions today. Miners compete for mining each block, as they receive a monetary reward (funded by transaction fees) for mining a block. To ensure immutability in the sequence of previously mined blocks, miners are obliged to include hash of the previous block and solve a computationally difficult cryptographic puzzle while mining. When a block is mined, the miner shares the block with the rest of the network by broadcasting it. A peer receiving a freshly mined block first verifies its authenticity, before appending the block to its local copy of the blockchain or relaying the block to other neighbors. 

\subsubsection{Block propagation delay and performance}
Blocks are broadcast in Bitcoin via flooding; when a peer receives a new block, it announces the hash of the block to all its neighbors via an \texttt{INV} message.  
Subsequently, neighbors who have not yet received the block respond with a \texttt{GETDATA} message requesting for the block, and the block is relayed to them.  
The process repeats until all the peers in the network have received the new block. 

The performance of Bitcoin, and other cryptocurrencies, is measured by their 
 (i) {\em throughput}, which is the average rate at which transactions are confirmed in the blockchain per second, 
 (ii) {\em confirmation latency}, the time taken such that the probability for removing an honest transaction from the blockchain becomes sufficiently small, and 
 (iii) {\em security}, the extent of adversarial peers the network can tolerate before the blockchain loses its immutability property~\cite{bagaria2019prism}.  
Cryptocurrencies today offer strong security guarantees, but are lacking in their throughput and confirmation latencies compared to mainstream payment systems.  
For example, Bitcoin promises its blockchain cannot be compromised as long as more than 50\% of the miners are honest.  
However, compared to the average throughput of $1700$ transactions per second in the Visa network, the average throughput in Bitcoin today is just 3--7 transactions per second, and the latency is 1 hour~\cite{croman2016scaling}.

A key factor affecting the throughput, confirmation latency and security is the propagation delay of blocks. 
If the propagation delay is too large, then there is a higher probability of mining of a block while another block at the same blockchain height is being propagated across the network---a phenomenon called forking~\cite{decker2013information}---reducing network throughput. 
The confirmation latency is also physically lower bounded by the propagation delay of the underlying p2p network~\cite{bagaria2019prism}. 
Furthermore, a large propagation delay can help an adversary to execute double spending and block-withholding attacks~\cite{sapirshtein2016optimal}.

\subsubsection{How the p2p topology impacts delay}
The dynamics of block propagation in Bitcoin has been empirically observed to follow a pattern similar to randomized rumor spreading in networks~\cite{decker2013information}.  
For instance, when a block is mined and broadcast, it first spreads exponentially fast to peers that are close to the source, before slowing down exponentially and reaching the remaining peers~\cite{decker2013information, karp2000randomized}. 
Prior works have extensively analyzed (both empirically, and theoretically) rumor spreading on different network topologies~\cite{doerr2011social, fountoulakis2010reliable}, and have shown that rumors spread substantially faster in certain topologies than others. 
Specifically, scale-free graphs spread rumors significantly faster (in sub-logarithmic time) than random graphs.  
Doerr et al.~\cite{doerr2011social} report that on social networks (e.g., the Twitter topology), rumors spread even faster than on scale-free graphs. 

While the rumor-spreading model is considerably simpler compared to the dynamics of block propagation in Bitcoin's network (e.g., it does not model heterogeneity in link latencies, or bandwidth) it illustrates the potential benefits of carefully designing the \pp topology. 
An optimal peer connection protocol should not only imbibe essential properties of a fast rumor-spreading network and take peer heterogeneity into account, but should also be implementable in a decentralized manner without introducing new vulnerabilities. 



\subsection{Problem Statement and Contributions} \label{s:probstmt}

We consider re-designing Bitcoin's \pp topology, to minimize the time taken by blocks to propagate over the network.
The topology is constructed using a fully decentralized protocol running at all the peers.  
A peer may choose to not follow protocol, or even act adversarially, but we assume there exist peers, whose aggregate compute power amounts to more than 50\% of the total compute power in the network, that are honest~\cite{nakamoto2019bitcoin}. 
Each honest peer seeks to connect to a set of neighboring peers, to minimize the time it takes for a block mined by the peer to reach a majority (e.g., 90\%) of the compute power in the network. 
Honest peers are also interested in receiving blocks mined by a majority of other peers 
as early as possible.
Our main contributions are as follows. 

\smallskip
\noindent 
{\bf Fundamental bounds on delay.}  
We present a theoretical model for analyzing block propagation delay in Bitcoin, that explicitly models heterogeneity in the communication latencies between peers. 
Our model is based on a line of work in the networking systems, which has proposed that latencies between hosts on the Internet can be accurately predicted by embedding the hosts on to a metric space~\cite{dabek2004vivaldi}. 
With this model, we show that inter-connecting peers randomly leads to propagation delays that are logarithmically worse compared to the underlying point-to-point latencies between peers  (\S\ref{s:alg:random}). 
Conversely, we also show that a topology in which peers choose neighbors with whom they have a small round-trip-time latency, provides asymptotically the best possible propagation delay (\S\ref{s:theoretical optimum}).  

\smallskip
\noindent
{\bf Optimal algorithm.}
We propose \namens, a decentralized neighbor-selection protocol, that adaptively decides which neighbors to connect to purely based on the interactions between a peer and its neighbors (\S\ref{s:design}). 
\name is motivated by the classical multi-armed bandit problem~\cite{auer2002using}
, in which an agent---faced with a decision to choose one among many options with a priori unknown rewards---adaptively tries the different options and zeros-in on the best choice. 
A core tenet of algorithms for solving the multi-armed bandit problem is balancing exploration (trying out a previously unexplored option) with exploitation (choosing an option that has already been tried before). 
In our case, each peer is an agent that is faced with choosing the best set of neighbors, among different choices for neighbors. 
Interpreting the \pp topology design problem as an instance of multi-armed bandit problem is a key novelty of our paper and, to our best knowledge, has not been proposed before. 
Further, our experiments show that the topology that is adaptively learned by \name has striking statistical similarities to the theoretically optimal topology (\S\ref{s:theoretical optimum}).   
In addition to minimizing propagation latency, \name is attractive also for the following reasons:
\begin{itemize}
\item It is lightweight.
\item It is compatible with the self-interests of peers---each peer selfishly tries to select the best neighbors for itself.
\item It supports incremental deployment: peers following \name would see improvements in how quickly they can send or receive blocks, compared to those that do not follow \namens.
\item It is robust against adversarial actions: a \name peer does not need to know much about a candidate neighbor (e.g., its  geographical location, or the round-trip latency to the neighbor) to decide whether to connect to it.
\item It incentivizes peers to relay blocks promptly. 
\item It is naturally adaptive to varying hash-power. Each node tries to optimize its distance from an average block source, rather than from an average node.
\end{itemize}     



%% file: model.tex
\section{System Model} \label{sec:model}

\subsection{Network Model} 

We model Bitcoin's \pp network as an undirected graph $G(V, E)$, where $V$ is the set of nodes, and $E$ is the set of edges, or links, between the nodes.    
A node refers to a Bitcoin server (e.g., a miner), that can accept incoming TCP connection requests from other servers and clients. 
Clients on Bitcoin are end-devices that are not able to accept incoming TCP connection requests (e.g., because they are behind a NAT).  
Once a TCP connection has been established between two nodes, communication can happen in both directions. 
We focus in this work, on Bitcoin servers as they form the core of the \pp overlay---servers tend to be always on, and the time taken for a block to propagate is largely affected by the interconnection network between the servers. 
We focus on minimizing the latency of propagating blocks, not transactions, in this work (we define the objective formally in \S\ref{s:performancemetric}).
It is well known that transaction throughput and confirmation latencies in Bitcoin are directly correlated with block propagation times~\cite{bagaria2019prism}. 
Moreover, a \pp network that is optimized for rapidly broadcasting blocks would also minimize transaction broadcast time, as clients are likely to connect to well-connected server nodes (e.g., using~\cite{bitnodescom}).    
However, our protocol is general, and can readily be adapted to optimize transaction propagation times as well. 

For any two nodes $u, v \in V$, we assume the latency of sending a block from $u$ to $v$ or from $v$ to $u$, via a TCP connection between $u$ and $v$, is a constant $\delta_{(u,v)} \geq 0$. 
The latency here includes transmission delay, in-network (propagation, queueing etc.) delays and protocol-specific message exchange overheads (e.g., \texttt{inv}, \texttt{getdata} exchange in Bitcoin) while sending a block. 
$\delta_{(u,v)}$, for a pair of nodes $u, v$, depends on various factors: the size of each block, the Internet access bandwidths at $u$ and $v$, the physical distance between the nodes, and the extent of congestion in the network.  
We assume these factors are slowly varying compared to the timescale of our algorithm. 
Each node $v \in V$ also spends a fixed amount of time $\Delta_v$, for cryptographically verifying the authenticity of a block it receives. 
$\Delta_v$ varies between nodes depending on their processing power. 
The fraction of hash power a node $v$ has, relative to the total hash power of the network, is denoted by $f_v$. 

We assume blocks are periodically generated (e.g., once every 10 minutes) and broadcast over the network. 
The probability that a node $v$ generates the block in a round is proportional to its hash power $f_v$. 
When a node $u$ mines a block, or receives a block from a neighbor, it immediately starts relaying the block to each neighbor $v$, taking a time $\delta_{(u,v)}$ to finish relaying. 
For simplicity
we also assume that the connection updates
execute synchronously at all the nodes, immediately after a block is broadcast on the network. 


At any time, each node maintains $d_\mathrm{out}=8$ outgoing connections, and has $d_\mathrm{in} \leq 20$ incoming connections. 
In practice, Bitcoin nodes can have up to 8 outgoing and 125 incoming connections~\cite{miller2015discovering}.  
To discover peers in the network, Bitcoin nodes also maintain a local database called \texttt{addrMan}, which they regularly update by exchanging messages to neighbors.  
A bootstrapping server provides with a list of addresses for a freshly joining peer. 
However, we assume each node know the IP addresses of all other nodes. 

\subsection{Performance Metrics} \label{s:performancemetric}

For each $v \in V$, we compute the minimum overall delay $\lambda_{v}$ it takes for a block mined and broadcast by $v$ to reach nodes totalling to at least 90\% of the hash power in the network. 
The objective for each $v \in V$ is to choose neighbors such that $\lambda_v$ is minimized. 
By symmetry, this objective would equivalently also minimize the time taken by blocks mined by a majority of other nodes to reach $v$.

%% file: baseline.tex
\section{Baseline Algorithms} \label{s:baseline}

\subsection{Random} \label{s:alg:random}

\begin{figure}[t]
    \centering
    \includegraphics[width=0.48\textwidth]{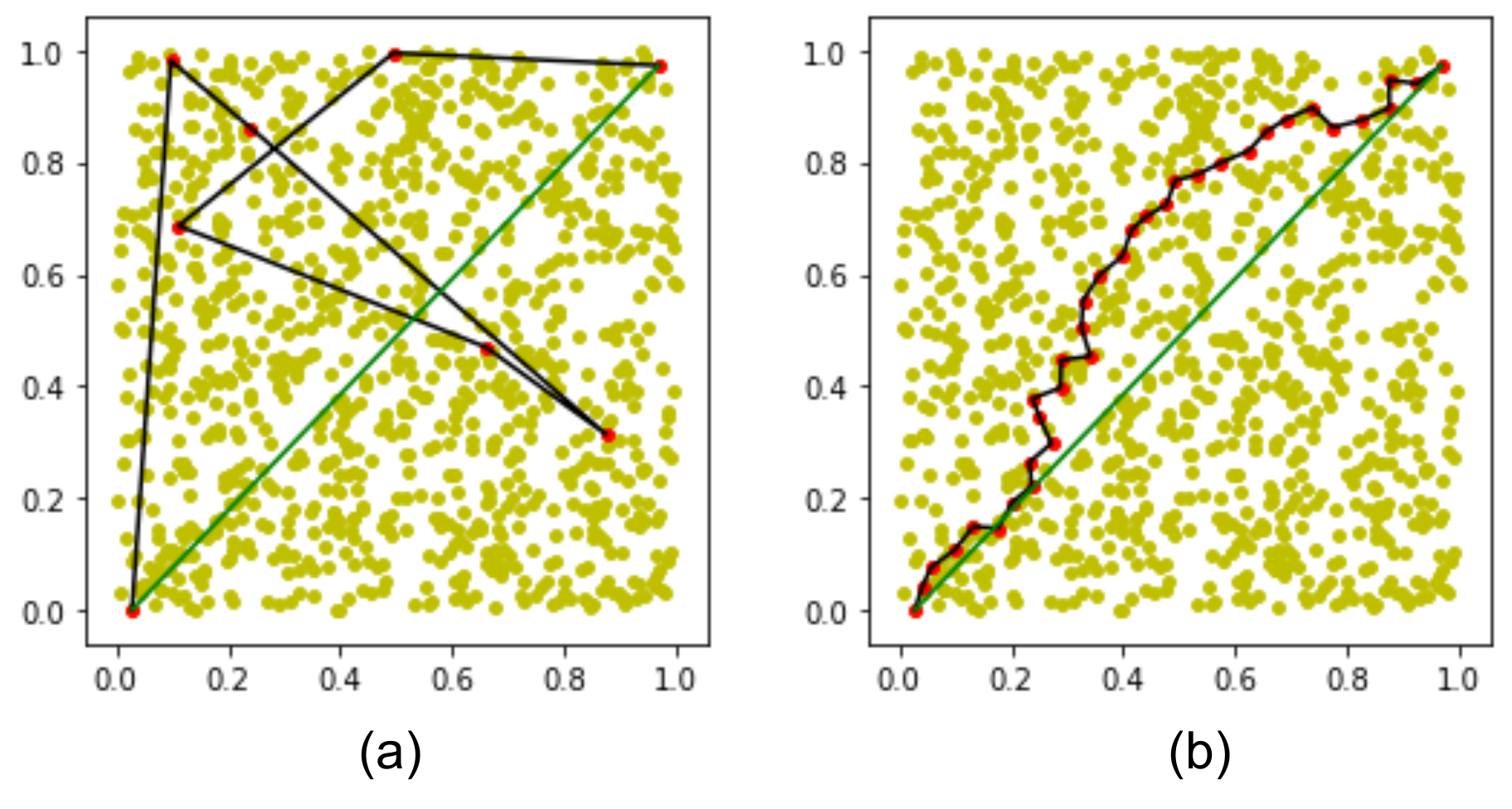}
        \vspace{-5mm}
    \caption{Example of 1000 nodes embedded randomly within a unit-square. (a) If nodes are interconnected according to a random topology, the shortest path between two points can be much longer than the Euclidean distance between the points. (b) If nodes are interconnected using a carefully designed topology (e.g., a geometric graph; see \S\ref{s:theoretical optimum}), significantly better paths, with length close to the Euclidean distance, are possible. }
    \label{fig:metricspace}
    \vspace{-5mm}
\end{figure}

The random connection policy is a simple algorithm that is widely deployed in many cryptocurrency systems today. 
In this algorithm, a node maintains a list of IP addresses of a small number of nodes that are currently active in the network. 
Initially a bootstrapping server provides the node with such a list; subsequently the list is updated (i.e., new addresses are added, while stale ones are removed) by gossiping any changes in the set of neighbors for each node, over the network. 
Intuitively, if connections are formed randomly on a world-wide network, then any path---and in particular the shortest path---between two nodes $u, v$ would likely pass through intermediate nodes that are not located close to the shortest geographical route (\ie the geodesic) connecting $u$ and $v$. 
Such less-than-direct paths would prolong the propagation delays of blocks sent on the network.  
Moreover, even with queueing delays on the Internet, we can show that a random topology leads to paths with latencies significantly larger than those of paths on optimal topologies. 
Based on extensive measurement studies, prior works~\cite{dabek2004vivaldi} have empirically shown that endhosts on the Internet can be {\em embedded} on a high-dimensional metric space (e.g., $\mathbb{R}^5$) such that the metric distance between any two endhosts accurately predicts the communication latency between the hosts. 
However, 
the paths on a randomly connected network are unlikely to remain close to the geodesic shortest route between hosts, on the embedded high-dimensional space. 

\smallskip
\noindent 
{\bf Example.}
To illustrate this, consider an example of a network embedded in the unit square $[0, 1] \times [0, 1]$, as shown in Figure~\ref{fig:metricspace}. 
The green points within the square are drawn uniformly randomly and represent the nodes in the network.
The Euclidean distance $||u-v||_2$ between any two nodes $u, v$ is the one-way latency of sending a message (e.g., a transaction, or a block) from $u$ to $v$ or vice-versa.\footnote{From the literature and results on metric-embedding of Internet hosts, we assume that message latency from $u$ to $v$ is equal to the latency from $v$ to $u$.}
Now, consider connecting each node in the unit-square randomly to 3 other nodes.
Figure~\ref{fig:metricspace}(a) shows the shortest path on this topology, between two nodes $a$ and $b$ that are closest to the bottom-left and top-right corner of the square.  
However, due to the meandering nature of paths in a random topology, the latency between $a$ and $b$ is much greater than the point-to-point latency $||a - b||_2$ between them. 
In contrast, a geometric graph topology (to be discussed shortly in \S\ref{s:theoretical optimum}) has a shortest path between $a$ and $b$ that is much closer to the geodesic shortest path (straight line between $a$ and $b$), as shown in Figure~\ref{fig:metricspace}(b).
We formally show the suboptimality of the random topology next.

\smallskip
\noindent 
{\bf Suboptimality of the random algorithm.}
Let $[0, 1]^d$ be the $d$-dimensional hypercube ($d \geq 2$), equipped with the Euclidean metric, and let $V = \{x_1,x_2,\ldots,x_n\}$ denote the nodes in the network. 
To model the point-to-point latencies between different pairs of nodes, \ie the latency between pairs of nodes if they are directly connected to each other, we consider an embedding of $V$ on to $[0,1]^d$, in which each node $x_i$ is mapped to a point $X_i$ chosen uniformly randomly over $[0, 1]^d$.
The point-to-point latency between any two nodes $x_i, x_j \in V$, is then simply $||X_i - X_j||_2$.   

Next, to model random connections between nodes, for each pair of nodes $x_i, x_j$ we let $x_i$ and $x_j$ have a link between them with probability $p$, independent of other links.    
Equivalently, we can consider each pair of points $X_i, X_j$ on the embedded space to have a link between them with probability $p$, independent of other links.     
The resulting random graph of points $\{X_1, X_2, \ldots, X_n \}$ is denoted by $\tilde{G}(\tilde{V}, \tilde{E})$. 
The network latency $\mathrm{dist}(i,j)$ between any two points $X_i, X_j$ is the time taken for a message broadcast by node $i$ (resp. node $j$) to reach node $j$ (resp. node $i$). 
This is computed as the total weight of edges on the shortest path between $X_i$ and $X_j$ on $\tilde{G}$, where the weight of each edge $(X_u, X_v) \in \tilde{E}$ is given by $||X_u - X_v ||_2$. 
Clearly, the maximum point-to-point latency between any two points is bounded by $\sqrt{d}$, which is the Euclidean distance between the diagonal points $[0, 0, \ldots, 0]$ and $[1, 1, \ldots, 1]$ on the hypercube. 
However, due to the random nature of the graph, the typical network latency between any two nodes $i, j$ can be a logarithmic factor worse as shown by the following Theorem.  
\begin{thm}[\cite{frieze2019traveling}] \label{thm: random}
For any pair of nodes $x_i, x_j \in V$ and $p \leq c \log n/n$, where $ c = c(n) = O(1)$, we have 
\begin{align}
\mathrm{dist}(i, j) \geq \frac{(\log n)^{1-\frac{1}{d}}}{8d^{3/2}e^d (\log \log n)^2 c^{1/d}} ||X_i - X_j||_2, \label{eq: thm 1}
\end{align}
with probability $1 - o(1)$. 
\end{thm}
(Proof in Freize et al.~\cite{frieze2019traveling}.)

In Theorem~\ref{thm: random} above, $p \leq c \log n / n$ connotes a small average degree of $c \log n$ per node in the network. 
The latency bound in Equation~\eqref{eq: thm 1} holds asymptotically almost surely for any pair of nodes $i, j$ because of our assumption that each node is embedded on to a random point in the hypercube.\footnote{This leads to any two points $X_i, X_j$ being "well-separated" on the hypercube with high probability.}   
In reality, while connection patterns across nodes can change randomly with time, the point-to-point latencies between nodes may not vary significantly. 
Nevertheless, for nodes that are not too close to each other, Theorem~\ref{thm: random} suggests that the latency between them on a random network must be logarithmically worse.

\subsection{Connecting Based on Geography} \label{s:geography}


A key reason the randomly formed topology suffers from suboptimal path delays (Theorem~\ref{thm: random}), is due to a lack of sufficient connectivity between nodes that are in close proximity (\ie have small delay) to each other in the hypercube.   
If the size of the network is large, each node is likely to choose neighbors that are all far away, as the number of distant nodes is much greater than the number of nearby nodes. 
Therefore, even if a message reaches the general vicinity of node fast, it likely needs to spend a disproportionate amount of time to actually reach the node, due to the lack of any direct, low-delay paths.  
To ensure good connectivity in this ``last mile'', it is desirable for nodes to connect not only to nodes that are far away, but also those close by.   

In practice, it is difficult for nodes to a priori know the round-trip-times to other nodes, without actually connecting to them first. 
However, recent work~\cite{bieri2019simulating} has proposed using the geographical location of a node---which can be estimated based on it's IP address~\cite{hu2012towards}---as a proxy for predicting whether the connection latency to the node is likely to be large or small.
If the geographical locations of nodes are known, then a natural method to improve the random protocol, is to select a few neighbors among those that are geographically close, and then choose the rest of the neighbors randomly. 
For instance, if we cluster nodes according to the continents they are from, then a node located in North America can have four neighbors that are also in North America, and four other neighbors from other continents (e.g., Asia, Europe). 

In our evaluations (\S\ref{s:eval}), we show that the above protocol, does indeed perform better compared to the random protocol.    
However, the question remains whether this protocol can be improved even further. 
For instance, 
we have clustered nodes based on the continent in which they are located, but it is unclear if a different way of grouping nodes would have fared better.    
We also 
assign 50\% of a node's connections to in-cluster nodes, and the remaining 50\% to nodes outside the cluster. 
The optimal balance between the number of connections made within and outside of the cluster, is again unclear. 
In practice a node may be malicious and try to spoof it's true geographical location (e.g., via proxies, or VPN), which can also significantly degrade the utility of the algorithm.   
Lastly, the assumption that the geographic distance to a node dictates the latency to it is only a coarse approximation
~\cite{hoiland2016measuring}. 


\subsection{Theoretical Optimum} \label{s:theoretical optimum}

To understand how much better an optimal topology can be, we consider a geometric graph in which two nodes are connected if the latency between them is less than a threshold $r$.  
Compared to the random topology, in which neighbors are selected completely agnostic of their delay or geography, the geometric graph represents the other extreme where  all neighbors are chosen to be within some small delay. 
Following the model for latency in which nodes are randomly embedded within a $d$-dimensional unit-hypercube (\S\ref{s:alg:random}), we can show that the shortest path distance between any two nodes is at most a constant factor larger than their Euclidean distance.  
\begin{thm}[\cite{solovey2018new, friedrich2013diameter}]
For a geometric graph with threshold $r = \Theta ((\log n / n)^{1/d})$, there exists a constant $\xi$ such that for any two nodes $x_i, x_j$ in the same connected component with $||X_i -  X_j||_2 = \omega(r)$, it holds that $\mathrm{dist}(i, j)$ is at most $\xi || X_{i} - X_{j}||_2 $ with probability $1 - o(1)$. 
\end{thm}
(Proof in Friedrich et al.~\cite{friedrich2013diameter})

The superior path delay of the geometric graph stems from nodes having a strong connectivity to other nodes in their local vicinity, which creates paths traversing closely to the geodesic between any two nodes (Figure~\ref{fig:metricspace}(b)).  
We note that the geometric graph is not the only construction with order-optimal path delay---a recent line of work
~\cite{dekker2012maximum, abraham2005metric} 
has proposed other efficient topology constructions also providing order-optimal path delays, for points embedded in a metric space. 
For example, in Chan et al.~\cite{chan2015new}, the authors propose a decentralized algorithm for constructing a low-stretch spanner where the graph distance between any two nodes is at most a constant factor worse than their Euclidean distance. 
Their algorithm is also robust against node faults.

The metric embedding model for node latencies, discussed in \S\ref{s:alg:random}, \S\ref{s:geography} and the present section, is useful as a simple, tractable theoretical model for analyzing competing topology constructions.  
While the model captures first-order differences in point-to-point latencies across nodes, blocks in the Bitcoin network also suffer from delays due to transmission (if the available bandwidth is small, relative to the block size), and block validation. 
Measurement studies on the Bitcoin network, report a wide skew in these delays across different nodes; e.g., in one study~\cite{croman2016scaling} conducted in 2015, the bandwidth of Bitcoin server nodes was found to vary from 3 Mbps to 186 Mbps.  
Bitcoin's block size has also varied over the years, from 87 KB in 2012 to around 1 MB today.
These numbers are likely to change as nodes continuously invest in better network infrastructure, compute and storage hardware,
and as the Bitcoin community introduces higher-level protocol changes
. 
To optimize the \pp topology in an evolving landscape, it is desirable for a neighborhood-selection protocol to be adaptive to changes, while at the same time have behavioral similarities to optimal topology constructions such as the geometric graph.  
\name is such a protocol; we discuss it next. 


%% file: design.tex
\section{\name} \label{s:design}

\name is a decentralized algorithm that adaptively {\em learns} to form optimal peer connections, purely based on a node's interactions with its neighbors. 
Unlike hand-crafted protocols which often require extensive manual tuning to optimize protocol parameters for individual blockchain networks, \name is a flexible learning algorithm that automatically finds the best topology for any network setting.
In \namens, nodes continuously monitor the promptness of block delivery from each of their neighboring nodes, and decide whether to retain their neighbors or explore connecting to other potentially better-connected peers. 
Since each node tries to locally find the best set of neighbors it can connect to, \name naturally  benefits the self-interests of peers. 
In \S\ref{s:eval} we show through extensive experimental evaluations, that our protocol also globally optimizes block propagation delays under diverse settings. 

\subsection{Algorithm Overview}


\name operates on top of existing block distribution protocols, and does not change the format of blocks or the gossip protocols used for broadcasting them. 
Instead it simply decides what is the best set of neighbors for a node to connect to, for a given block size and gossip protocol. 
Starting from an arbitrary initial set of neighbors (e.g, obtained randomly from a bootstrapping server), a node in \name periodically evaluates its current set of neighbors, to decide which neighbors offer the fastest connectivity to the rest of the network. 
Connections to those neighbors providing a good connectivity are retained, while the rest are disconnected. 
Additionally, \name also periodically connects to a small number of random nodes, as a means to discover previously unknown but potentially well-connected peers. 

In \namens, a neighbor is evaluated purely based on timestamp measurements of when blocks, or advertisements for blocks, were received from the neighbor.\footnote{Our protocol is general, and can also be used with timestamp measurements of transactions received from neighbors.} 
Based on these measurements, a real-valued {\em score} is computed and assigned to each neighbor, which is then used to decide which subset of neighbors to retain.  
Using block reception times to score neighbors, is a key novelty in \name and has several advantages compared to algorithms discussed in \S\ref{s:baseline}.  
A node is identified only by its IP address, and not based on auxiliary information such as its geographical location. 
This makes our algorithm robust to geo-spoofing attacks. 
Moreover, by explicitly using block arrival times for scoring neighbors, \name automatically takes into account heterogeneities, such as variations in link latencies across geographically separated nodes, and variations in hash power. 
The resulting topology therefore, has a good connectivity to nodes with high hash power, rather than good connectivity in a simpler graph theoretic sense (e.g., low diameter). 

\begin{algorithm}[t]
\DontPrintSemicolon
\SetKwInOut{Input}{input}\SetKwInOut{Output}{output}
\Input{
neighbors $\Gamma_v$, outgoing neighbors $\Gamma_v^o$, set of blocks $B$ mined during the round, and observation set $\mathcal{O}_{v}$ 
}
\Output{updated set of outgoing neighbors $\Gamma_v^o$ for next round}
\tcc{Score each neighbor based on measurements collected in $\mathbb{O}_v$ using a scoring algorithm (see \S\ref{s:independent}, \S\ref{s:groupscoring} for different scoring methods)}
score($u$) $\leftarrow$ \textsc{ScoringAlgorithm}$(\mathbb{O}_v)$, for each ngbr. $u \in \Gamma_v^o$ \;
\tcc{Retain subset of $d_v$ neighbors with best score}
$\Gamma_v^o \leftarrow \{ u \in \Gamma_v^o: \text{score}(u) \in \text{ best } d_v \text{ scores of nodes in } \Gamma_v^o  \}$ \;
\tcc{Additionally connect to $e_v$  random peers for exploration}
$\Gamma_v^o \leftarrow \Gamma_v^o \cup (e_v \text{ randomly chosen neighbors from } V) $
\caption{{\sc \namens:} Algorithm template for updating neighbors of node $v$ after each round.}
\label{algo:highlevel}
\end{algorithm}

To simplify our exposition, we present \name (Algorithm~\ref{algo:highlevel}) under the network model of \S\ref{sec:model}. 
The algorithm proceeds in rounds, where each round spans the time taken to mine and broadcast $K$ unique blocks $B= \{b_1, b_2, \ldots, b_K \}$ over the \pp network. 
For a node $v \in V$, let $\Gamma_v$ denote the set of $v$'s neighbors in $G$, and let $\Gamma^o_v \subseteq \Gamma_v$ denote $v$'s outgoing neighbors.
When a block $b \in B$ is broadcast during a round, we let $t_{u,v}^b$ be the local time at $v$ when $b$ was received from neighbor $u \in \Gamma_v$. 
We set $t_{u,v}^b = \infty$ if block $b$ was never relayed to $v$ by $u$. 
During a round, each node $v$ collects information about when each block was received from its neighbors in the form of an observation set $\mathbb{O}_v = \{ (b, u, t_{u,v}^b) : b \in B, u \in \Gamma_v \}$. 
Note that it is possible for $v$ to hear about a block for the first time from a non-outgoing neighbor. 
The tuples collected in $\mathbb{O}_v$ allows \name to rate how quickly a neighbor relays blocks relative to other neighbors, and retain connections to the best subset of $d_v$ (e.g., $d_v = 6$) neighbors at the end of each round.   
Neighbors are evaluated using a scoring function, which estimates the maximum delay taken by a neighbor to forward 90\% of blocks to the node $v$. 
In addition, \name also connects to a small number $e_v$ (e.g., $e_v = 2$) of random peers during each round, for discovering previously unknown peers with good connectivity.  

We propose two different scoring methods, depending on whether each neighbor is scored individually (\S\ref{s:independent}), or groups of neighbors are jointly assigned a score (\S\ref{s:groupscoring}).  
In the latter case, the score is an estimate of the maximum delay taken by the group of neighbors as a whole to forward 90\% of blocks to $v$. 
The joint scoring of different neighbor groups is better suited to our objective, since each peer ultimately just desires to receive blocks as fast as possible, regardless of which specific neighbor forwards the block. 
Certain blocks may be relayed fast by some neighbors, while other blocks are relayed fast by the remaining neighbors---as long as, together, the set of neighboring nodes result in a quick delivery of a majority of blocks, it is beneficial for the node.  
However, accurately evaluating scores for all possible groups of neighbors is computationally expensive, and therefore we propose faster approximate methods.  
In our evaluations (\S\ref{s:eval}), we find both the independent scoring and approximate joint scoring methods to be competitive. 

\subsection{Scoring Each Neighbor Individually}
\label{s:independent}


\subsubsection{Vanilla Scoring} \label{s:vanillascore}
We first consider a simple scoring method, called \textsc{VanillaScoring}, to illustrate how for each neighbor $u \in \Gamma_v^o$ of a node $v$, the timestamps $t_{u,v}^b$ of blocks $b \in B$ broadcast during a round can be used to estimate $u$'s score. 
Recall that a timestamp $t_{u,v}^b$ recorded in the observation set $\mathbb{O}_v$ of a node $v$ corresponds to the local wall-clock time when the block $b$ was received at $v$. 
In order to judge how well a neighbor is connected to the rest of the network, it is desirable to know the {\em relative} time between when a block was mined, and when it was delivered by the neighbor (\ie the propagation delay). 
However, as it is difficult for a node $v$ to know the precise time when a block was mined, we use the relative time differences between when a block is forwarded by different neighbors, as a proxy for the propagation delay.  
For a block $b$, the first time it was received by $v$ from some neighbor is at time $t_v^b := \min_{u\in \Gamma_v} t_{u,v}^b$.  
The timestamps in $\mathbb{O}_v$ are then revised relative to times blocks were first received by $v$, and a time-normalized observation set $\tilde{\mathbb{O}}_v$ is computed as 
\begin{align}
\tilde{\mathbb{O}}_v = \{ (b, u, t_{u,v}^b - t_v^b): u \in \Gamma_v, b \in B  \}.
\end{align}
In \textsc{VanillaScoring}, the score for a neighbor $u \in \Gamma_v^o$ is simply computed as the 90th percentile of the multi-set of relative timestamps $\tilde{T}_{u, v} := (\tilde{t}: (b, u, \tilde{t}) \in \tilde{\mathbb{O}}_v)$ observed in a round. 
This scoring approach naturally reflects a node's preference to retain an outgoing neighbor from which it received transactions relatively earlier.
The lower the score for a neighbor, the higher is the preference for node $v$ to retain the neighbor in next round. 


\subsubsection{UCB Scoring}
\label{sec:ucb}

\begin{figure}[t]
    \centering
    \includegraphics[width=0.48\textwidth]{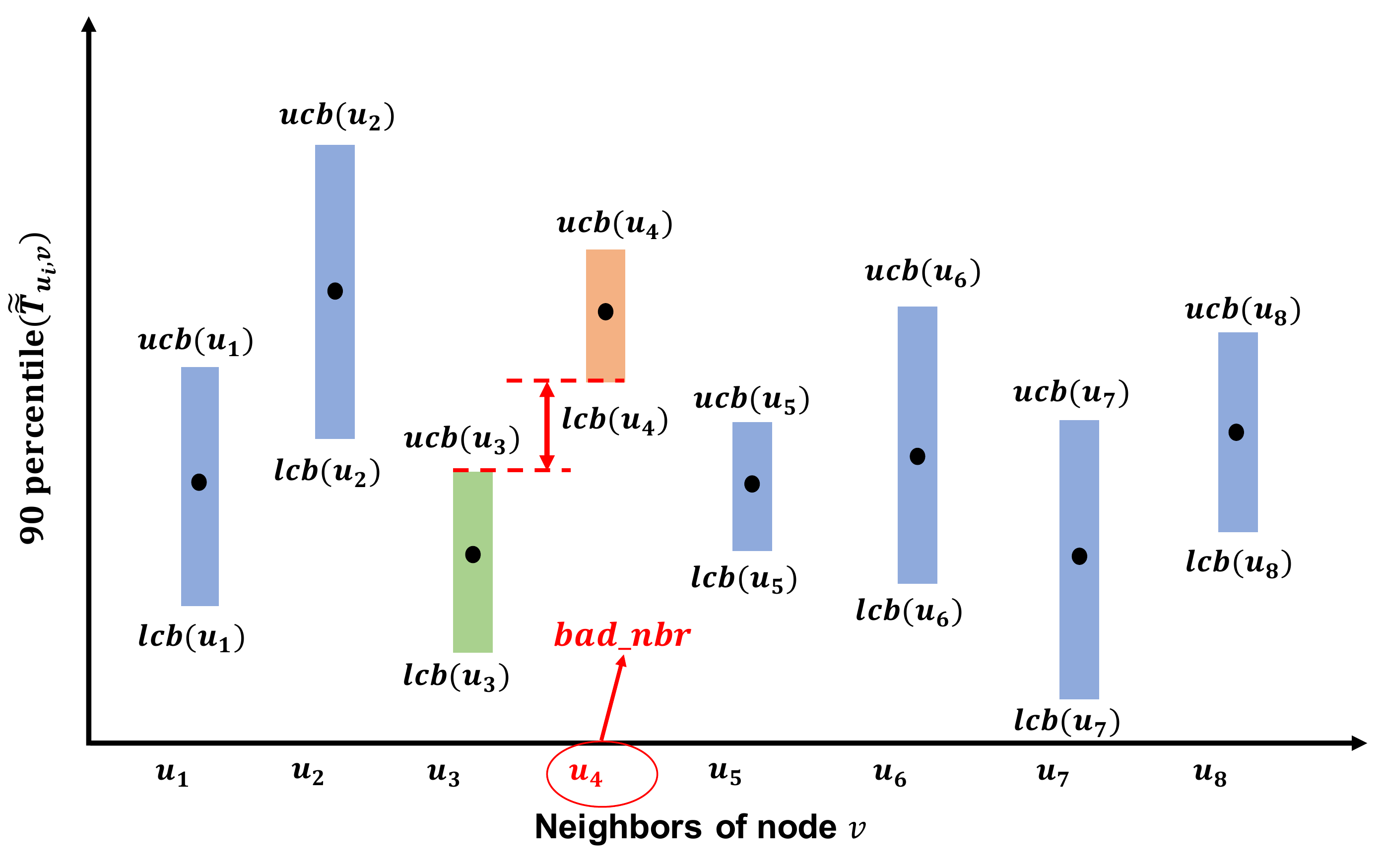}
    \caption{Observe that $\text{lcb}(u_{4}) \geq  \text{lcb}(u)$ $\forall u \in \Gamma_v^o $, $\text{ucb}(u_{3}) \leq \text{ucb}(u)$ $\forall u \in \Gamma_v^o $. Moreover,  $\text{lcb}(u_{4}) > \text{ucb}(u_{3})$ and so the neighbor to be removed ($bad\_nbr$) is $u_{4}$.}
    \label{fig:ucb}
    \vspace{-5mm}
\end{figure}

In the \textsc{VanillaScoring} method, propagation delay estimates for individual neighbors (90th percentile of relative timestamp observations) are likely to be noisy if the number of blocks $|B|$ in a round is small. 
The noise here arises due to the randomness in which node mines a block each time (\S\ref{s: background}). 
While increasing $|B|$ by increasing the duration of each round improves accuracy of our estimates, it also slows down the overall convergence time of the algorithm.\footnote{We illustrate convergence of \name empirically in our experiments in \S\ref{s:eval}.}  
To improve the accuracy of the \textsc{VanillaScoring} estimates, without sacrificing on convergence time, we propose a second scoring method motivated by the Upper Confidence Bound (UCB) algorithm for multi-armed bandits~\cite{auer2002using}. 
In the \textsc{UCBScoring} method, a node maintains an estimate of propagation delay  for a neighbor, based on observed timestamps, and also computes lower and upper confidence bounds for it.    
If a neighbor has been connected to $v$ for longer than one round, then the estimates and confidence bounds for the neighbor are computed not only using the observations $\mathbb{O}_v$ made during the current round, but also using past observations available for the neighbor. 
For a neighbor $u\in \Gamma_v^o$, let $\tilde{T}_{u,v}(-i)$ denote the multi-set of relative timestamps obtained during a round $i$ rounds before the present round.  
Supposing node $u$ has been $v$'s neighbor for the past $r_{u,v}$ rounds. 
In the \textsc{UCBScoring} approach, we use a multi-set of relative timestamp observations $\dbtilde{T}_{u,v} = (\tilde{t}: (b, u, \tilde{t}) \in \cup_{i=0}^{-r_{u,v}} \tilde{T}_{u,v}(-i) \text{ such that } \tilde{t} < \infty )$ for a neighbor $u$.\footnote{Note that the union $\cup_{i=0}^{-r_{u,v}} \tilde{T}_{u,v}(-i)$ is a multi-set union.} 
The propagation delay for $u$ is estimated as the 90th percentile of $\dbtilde{T}_{u, v}$, 
and its
confidence bounds are computed as 
\begin{align}
\text{ucb}(u) = \mathtt{90percentile}(\dbtilde{T}_{u,v}) + c\sqrt{\frac{\log(|\dbtilde{T}_{u,v}|)}{2 \times |\dbtilde{T}_{u,v}|}} \label{eq:ucb} \\
\text{lcb}(u) = \mathtt{90percentile}(\dbtilde{T}_{u,v}) - c\sqrt{\frac{\log(|\dbtilde{T}_{u,v}|)}{2 \times |\dbtilde{T}_{u,v}|}}, \label{eq:lcb}
\end{align}
where ucb and lcb denote the upper and lower confidence bounds respectively~\cite{auer2002using}, and $\mathtt{90percentile}(\cdot)$ computes the 90th percentile of its argument. 
At the end of each round, in the \textsc{UCBScoring} approach we check whether $\max_{u\in\Gamma_v^o}(\text{lcb}(u)) > \min_{u\in\Gamma_v^o}(\text{ucb}(u))$, and if so, $v$ disconnects from the neighbor $\arg \max_{u\in\Gamma_v^o} (\text{lcb}(u))$ and connects to a random new neighbor instead; otherwise the current set of neighbors are retained for the next round. 
Figure~\ref{fig:ucb} shows an example of upper and lower confidence bounds for a set of eight neighbors. 
Node $u_4$ will be disconnected at the end of the round in this example. 
Updating the set of neighbors this way based on confidence intervals, avoids accidentally disconnecting from a well-connected neighbor that has a poor 90th percentile score due to randomness in mining and lack of sufficient measurement samples.

\begin{figure*}[!tbp]
  \centering
  \subfigure[]{\includegraphics[width=0.71\textwidth]{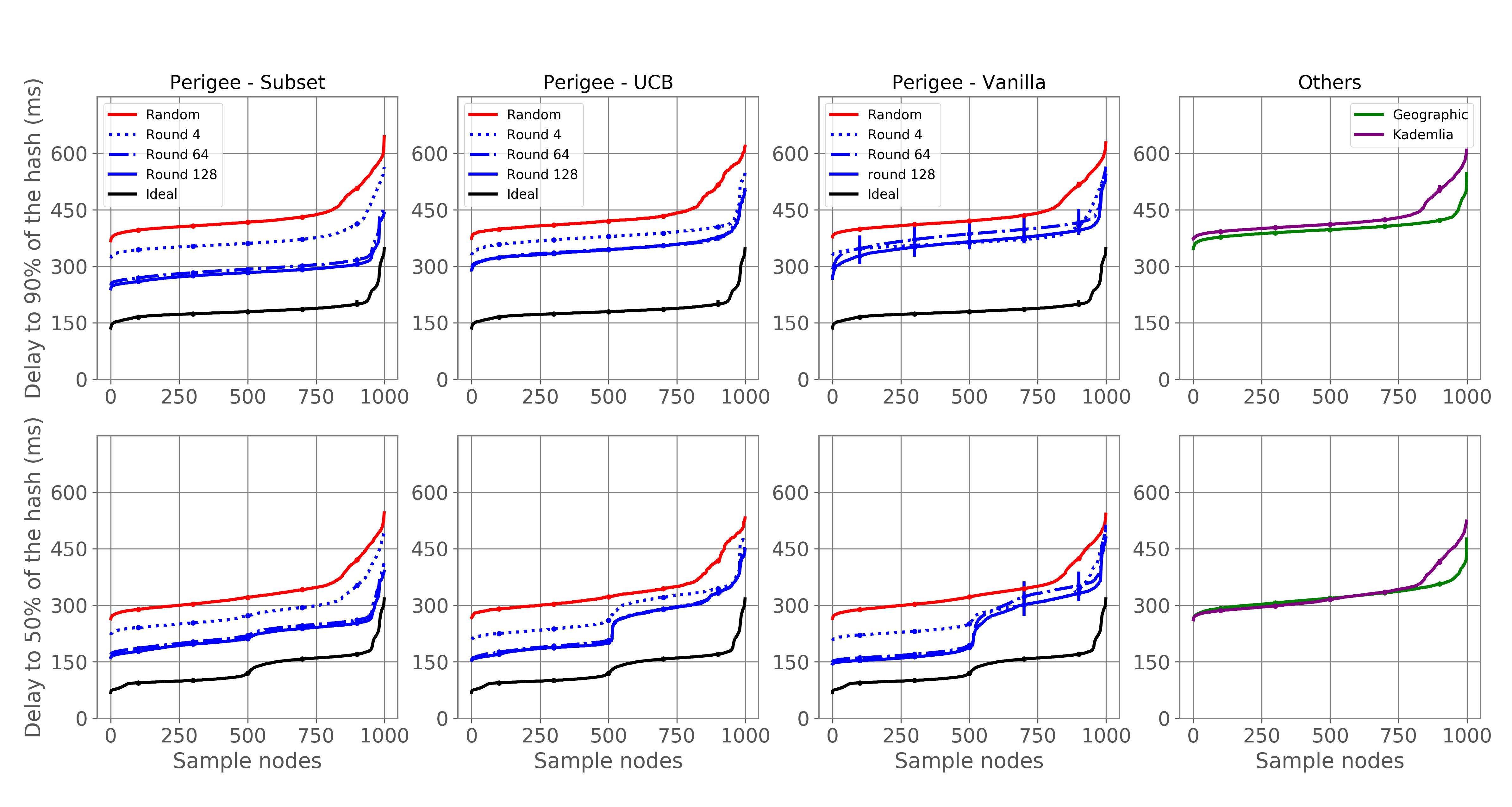}\label{fig:UniformHash}}
  \subfigure[]{\includegraphics[width=0.225\textwidth]{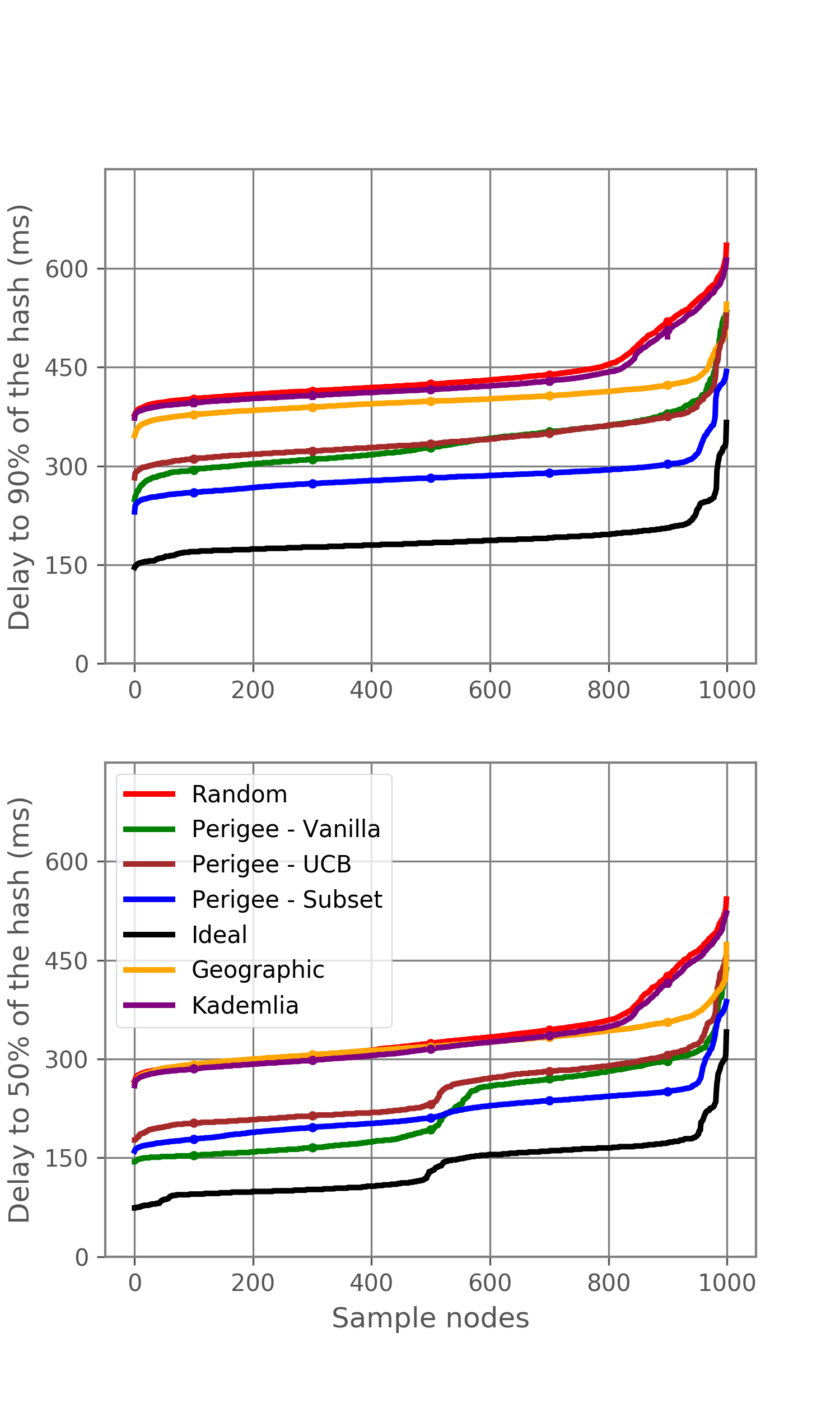}\label{fig:ExpDistributedHash}}
  \caption{Minimum delay to nodes totaling 90\% of network's hash power on random, geographic, \namens-Vanilla, \namens-Subset, \namens-UCB, Kademlia and the fullly-connected graph (denoted as ideal). (a) All nodes have the same hash power. (b) Nodes have a hash power drawn from an exponential distribution. 
  }
  \label{fig:Hash}
      \vspace{-3mm}
\end{figure*}

\subsection{Scoring Groups of Neighbors Jointly}
\label{s:groupscoring}

Next, we present an alternative scoring method, \textsc{SubsetScoring}, where scores are assigned  to each group $\gamma_v \subset \Gamma_v^o$ of neighbors (of a certain cardinality, e.g., 6) instead of to individual nodes. 
At the end of a round, the group of neighbors having the best score are retained and neighbors that are not part of this group are disconnected. 
As before, a small number of neighboring connections are made randomly in each round to encourage exploration. 

To avoid the computational overhead of exhaustively evaluating scores for all possible subsets of neighbors, we consider a simpler, but approximate, greedy approach in which the neighbors to be retained are selected one by one.  
First, the algorithm selects the neighbor $u_1 \in \Gamma_v^o$ having the best 90th percentile score in the relative timestamp observation multi-set $\tilde{T}_{u,v}$ (\S\ref{s:vanillascore}). 
If $k$ neighbors $u_1, u_2, \ldots, u_k$ have been selected, the $(k+1)$st neighbor is selected by first computing a transformed observation set $\dbtilde{O}_v(u_1,u_2,\ldots,u_k) = \{ (b, u, \min(\tilde{t}_{u,v}^b, \min_{1\leq i \leq k} \tilde{t}_{u_i, v}^b)): b \in B, u \in \Gamma_v^o \backslash \{u_1, u_2, \ldots, u_k\} \} $, followed by the multi-set of relative timestamps $\dbtilde{T}_{u,v}(u_1,\ldots,u_k) = (\dbtilde{t}: (b, u, \dbtilde{t}) \in \dbtilde{O}_v(u_1,\ldots, u_k))$ for each neighbor $u \in \Gamma_v^o \backslash \{u_1,\ldots, u_k \}$. 
The transformation essentially avoids penalizing nodes that do not have good connectivity to a certain part of the network, to which the neighbors already chosen have a good connectivity. 
The node $u \in \Gamma_v^o \backslash \{u_1,\ldots, u_k \}$ with the lowest $\mathtt{90percentile}(\dbtilde{T}_{u,v}(u_1,\ldots,u_k))$ value is finally selected as the $(k+1)$st choice. 
Thus, each time a neighbor is chosen whose connectivity to the rest of the network best complements the other neighbors selected thus far. 
As in \S\ref{s:independent}, once $(d_{v} - e_{v})$ neighbors are selected, node $v$ also randomly selects $e_v$ nodes as part of exploration. 
This set of $d_v$ nodes are $v$'s updated set of neighbors that it will monitor in the next round. 

%% file: eval.tex
\section{Evaluation} \label{s:eval}

We evaluate the performance of \namens, and compare it against the baseline algorithms of \S\ref{s:baseline}. Our experiments are based on a Python simulator we built following the network model of \S\ref{sec:model}.\footnote{Source code and datasets are available at \url{https://github.com/mori94/perigee}.} 
We describe the experimental setting in \S\ref{s:expt setup}. 
Following this, we evaluate \name on a variety of different network conditions (\S\ref{s:hashpower}--\S\ref{s:fastdistnet}).
\subsection{Experimental Setup}
\label{s:expt setup}

\begin{figure*}[!tbp]
  \centering
  \subfigure[]{\includegraphics[width=0.54\textwidth]{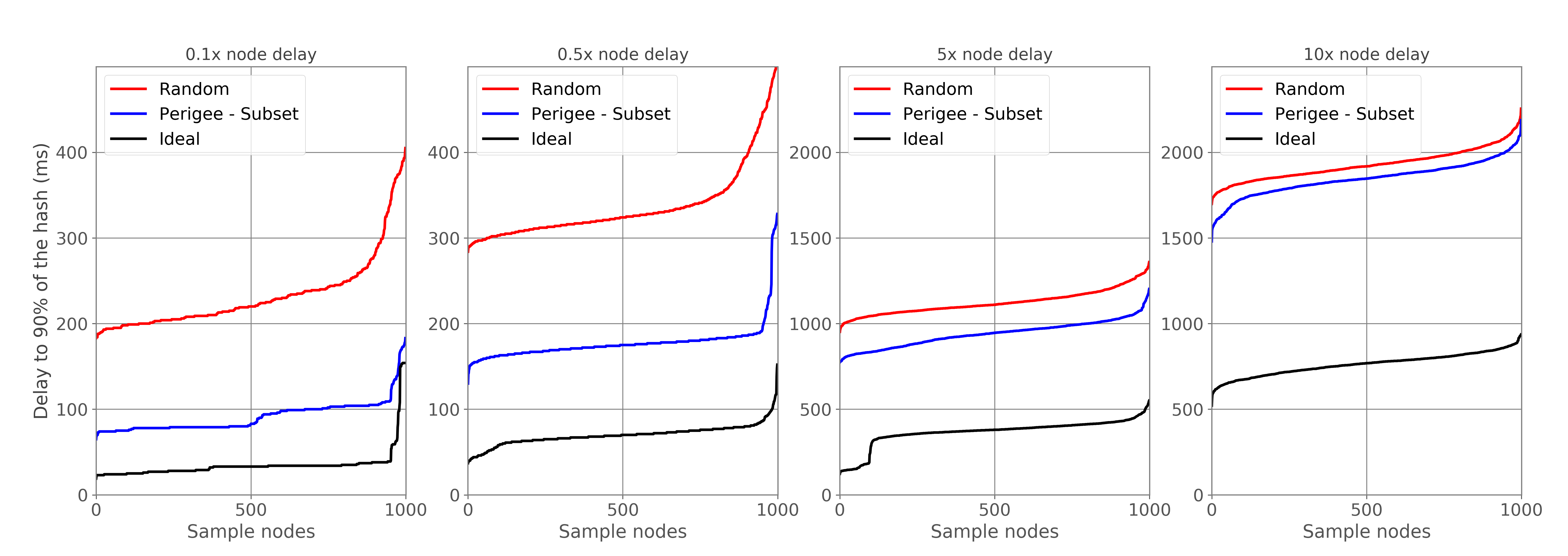}\label{fig:NodeDelay}}
    \subfigure[]{\includegraphics[width=0.21\textwidth]{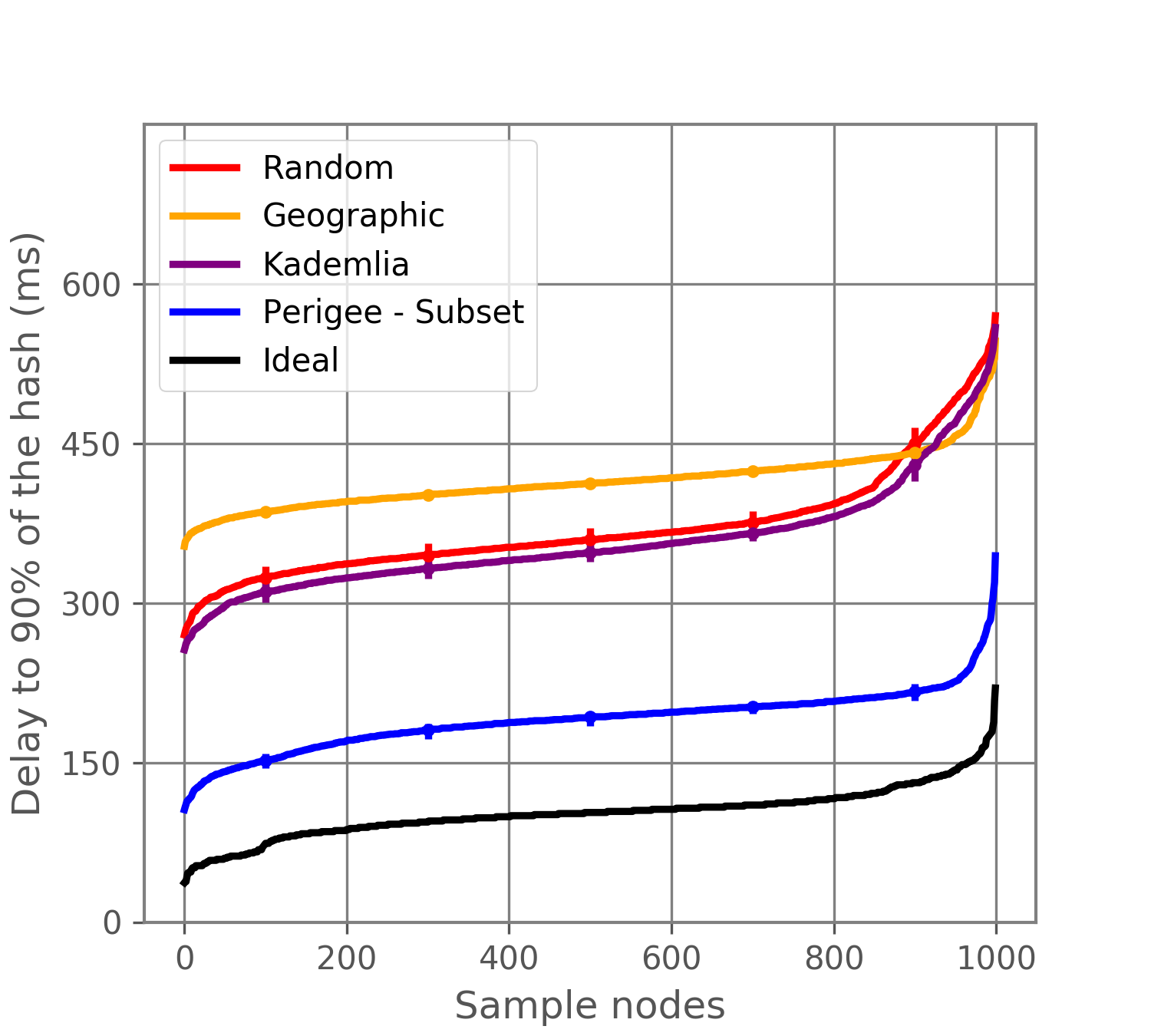}\label{fig:LowLatencySubgraph}}
  \subfigure[]{\includegraphics[width=0.21\textwidth]{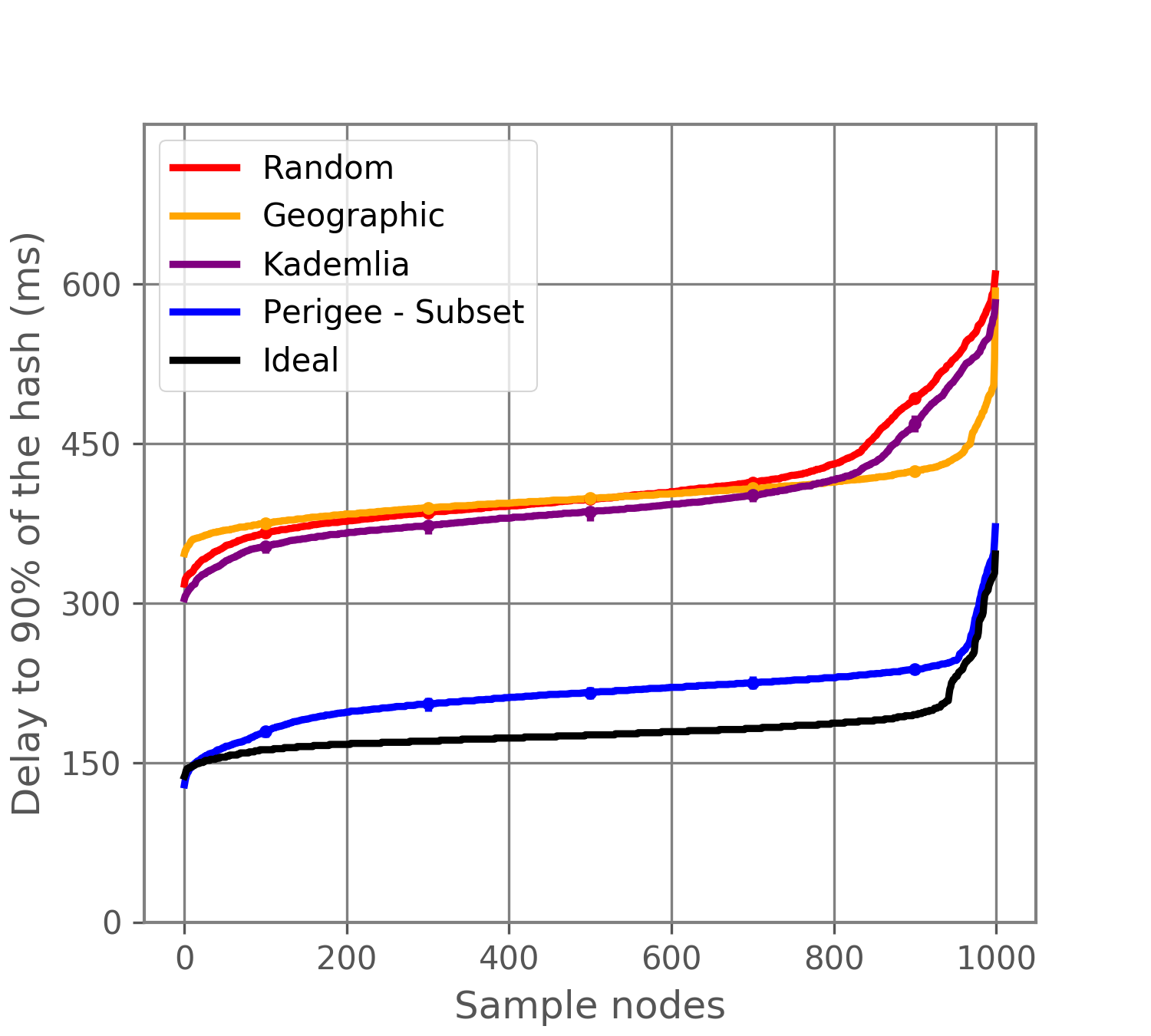}\label{fig:LowLatencyTree}}
  \caption{(a) Delay distributions for \name with 0.1$\times$, 0.5$\times$, 5$\times$ and 10$\times$ the default node delay. (b) Setting with a small number (10\%) of high hash power miners. (c) Performance in the presence of a low-latency block distribution network such as bloXroute.}
  \label{fig:NodeDelay}
   \vspace{-3mm}
\end{figure*}

\noindent 
{\bf Network setting.}
We retrieved a publicly available list of 9408 Bitcoin nodes~\cite{bitnodescom}, and use a randomly sampled subset of 1000 nodes from it, for all our experiments. 
The ``default'' setting for hash power of nodes, block validation times, link propagation delay, and block size in our experiments are described below.
In \S\ref{s:hashpower}, \S\ref{s:procdelay}, 
we consider a broader range of settings for each of hash power, block validation times 
respectively; in each case, while we explore different settings for one attribute, we fix the other attributes to their ``default'' setting unless specified otherwise. 
The default settings are as follows. \\
(1) {\em Hash power.} We assume hash power is distributed uniformly across all the nodes. \\
(2) {\em Propagation delay.} The dataset of Bitcoin nodes~\cite{bitnodescom} includes information about the geographical location of each node.  
Nodes are spread across seven geographic regions: North America, South America, Europe, Asia, Africa, China, and Oceania. 
We set the propagation latency between any two nodes according to their geographical locations, using the iPlane latency measurement dataset~\cite{madhyastha2006iplane, rohrer2019kadcast}. 
\\
(3) {\em Block size.} We assume block sizes are small, relative to the bandwidth available at the nodes. Hence the overall block broadcasting delay is dominated by the link propagation delays, and block validation delays, in the default setting.  \\
(4) {\em Block validation time.} Each node has a mean block processing time of 50 ms. \\
In addition, in \S\ref{s:fastdistnet} we consider a scenario where nodes have access to a high-speed block distribution network such as BloXroute~\cite{klarman2018bloxroute}. 
Each node creates 8 outgoing connections, and accepts up to 20 incoming connections.
If a node already has 20 incoming connections, any additional connection request is declined by the node.

\noindent 
{\bf Algorithms compared.} 
We implement \name under the scoring methods discussed in \S\ref{s:vanillascore}, \S\ref{sec:ucb}, \S\ref{s:groupscoring}, and name them \namens-Vanilla, \namens-UCB and \namens-Subset respectively.  
For \namens-Vanilla and \namens-Subset, we define a round such that $|B| = 100$ blocks are mined during each round; for \namens-UCB we use shorter rounds in which only one block ($|B| = 1$) is mined each round.  
In all of the \name variants, a node selects two neighbors randomly for exploration every round (\S\ref{s:design}).  
As baseline algorithms, we consider the random connection algorithm (\S\ref{s:alg:random}), geography-based connection algorithm (\S\ref{s:geography}) and a structured p2p topology based on Kademlia~\cite{rohrer2019kadcast}. 
For these baselines, we do not change the topology with each round. 
We also consider a topology in which each node is connected to all other nodes, to obtain a theoretical lower bound on block propagation times. 


\noindent 
{\bf Performance metric.}
For each node, we compute the time it takes for a block broadcast by the node to reach 50\% and 90\% of the hash power in the network.
We repeat each experiment three times using independently sampled link latencies, and plot the mean propagation times for different nodes in ascending order; we also show error bars at the 100th, 300th, 500th, 700th and 900th node.
Note that the nodes corresponding to the same $x$-coordinate value may not be the same node in the network. 

\subsection{Hash Power}
\label{s:hashpower}


We first consider the setting where all the attributes---hash power, link propagation latencies, block validation times, block size---take their default setting (\S\ref{s:expt setup}) and plot the results in Figure~\ref{fig:UniformHash}. 
The \namens-Subset and -UCB algorithm achieve around 33\% and 11\% lower delay respectively compared to random neighbor selection, indicating that switching neighbors based on their scores helps reduce the block propagation delays. 
Connecting based on node geography does help lower delay compared to random selection, however it is still 40\% worse than \namens-Subset at the 500th node. 
The Kademlia topology is slightly worse than even the geographic topology. 
While the 90-percentile delays in \name converge as the number of rounds increases, we observe the 50-percentile delays do not exhibit a similar monotonicity. 
This is because \name chooses neighbors only to optimize nodes' 90-percentile delays. 

Since \namens-Subset is slightly better than \namens-UCB or -Vanilla, for the reminder we have used \namens-Subset as the preferred scoring method. 
Next, we consider the same setting as above but where the hash power of the nodes are sampled from an exponential distribution (of mean 1), and normalized to 1 (Figure~\ref{fig:ExpDistributedHash}). 
The results show a similar performance pattern as in Figure~\ref{fig:UniformHash} with \namens-Subset being 33\% better than random.

\subsection{Processing Delay} 
\label{s:procdelay}

In Figure~\ref{fig:NodeDelay}, we vary the block validation time to 0.1$\times$, 0.5$\times$, 5$\times$ and 10$\times$ its default value. 
As shown in the Figure, for small values of node delay (0.1$\times$), \name finds a topology with delays at least 62\% better than random. 
However, as the node delay increases, \name approaches the random protocol's performance. 
This is expected, since with large processing delays the 90th percentile delay is dictated by the number of nodes on the shortest paths to nodes (\ie the diameter of the network). 
With node degree bounded by a constant, the diameter is lower-bounded by the logarithm in number of nodes, which is achieved by the random topology. 


\subsection{Fast Distribution Networks}
\label{s:fastdistnet}

The Bitcoin network is known to have a small number of mining pools that contribute to most of the hash power in the network. 
To simulate such a network, we randomly select 10\% of the nodes and assign them 90\% of the network's total hash power; we also set the link propagation latencies between the high-power miners to be much smaller than their default values. 
In this network, it is desirable for peers to be directly connected to at least one of the high-power miners. 
As shown by the results in Figure~\ref{fig:LowLatencySubgraph}, \name can exploit and explore the network to get much closer to the ideal delay in a fully-connected network compared to baselines. 
We also simulate fast block relay networks, by considering 100 nodes organized as a tree topology with low-propagation-latency links. 
The block validation delays for these 100 nodes are also set to be 10\% of their default value. 
Even here, as before, our results in Figure~\ref{fig:LowLatencyTree} show that \name can approach the fully-connected network baseline closely. 

\subsection{What does \name ~learn?}
\label{s:interpret}

\begin{figure}[!tbp]
  \centering
    \includegraphics[width=0.47\textwidth]{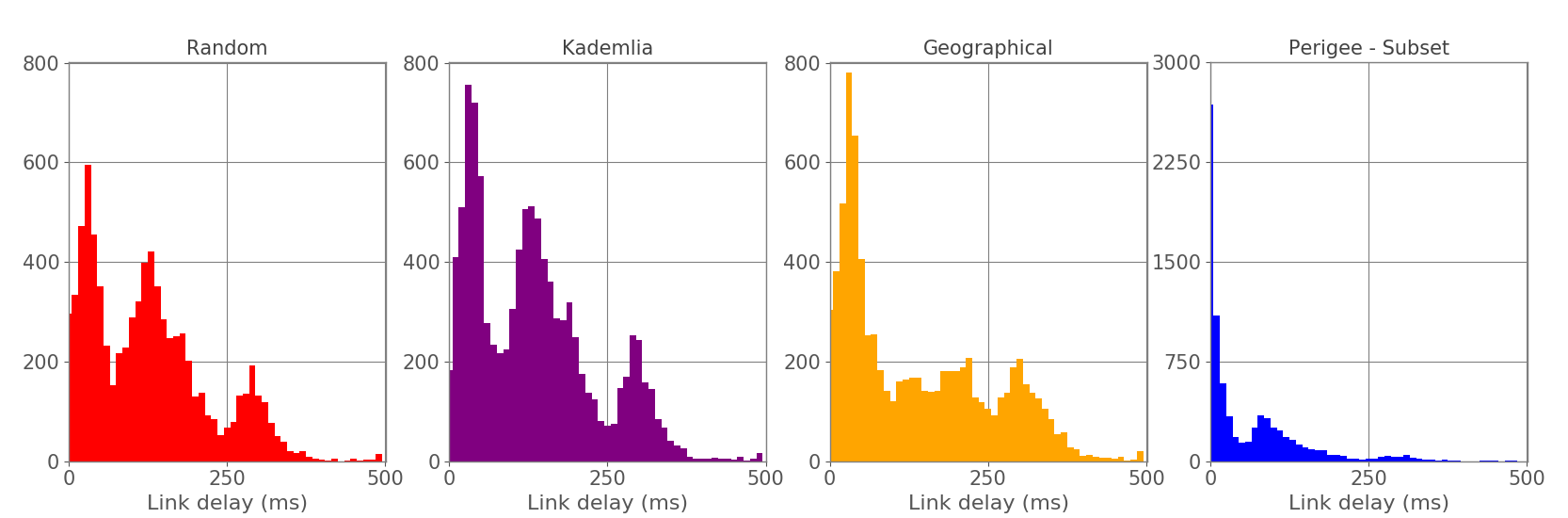}
  \caption{
  Histograms of the edge latencies in the \pp graph obtained after the execution of various algorithms under uniform hash power.}
  \label{fig:f2}
     \vspace{-3mm}
\end{figure}


In Figure~\ref{fig:f2} we observe that the distribution of edge latencies of the \pp network obtained in all the four algorithms are bimodal. 
The lower mode is mostly populated by intra-continental edges with smaller edge latencies whereas the upper mode is mostly populated by inter-continental edges with larger edge latencies. For \namens-subset, the latencies of bulk of the edges are populated around the lower mode. On the other hand, this is not the case in random and geometric. This implies that over the course of execution of \namens-subset algorithm, nodes learn to select those outgoing neighbors with which they have smaller edge latency.



%% file: conclusion.tex
\section{Discussion} \label{s:conclusion}

We have proposed \namens, an adaptive algorithm motivated by the multi-armed bandit problem, that finds efficient \pp topologies for reducing block propagation times in blockchain networks. 
While we have empirically illustrated the effectiveness of \namens, we believe our work is only a first step and important questions---both theoretical issues and practical considerations---need to be addressed for a more thorough understanding of the problem. 

Theoretically analysis of \namens, e.g., to study its convergence behavior and characterize its "regret" (how far it is from the "best" topology), is a crucial topic for future research. In \namens, one way to launch an Eclipse attack ~\cite{heilman2015eclipse} is for an adversary to provide blocks earlier than other nodes, thus gaining a peer's trust and dominating its neighborhood. 
The presence of random neighbors in \name provides some mitigation against this attack, a formal analysis of which is left for future work. 
Another dimension of analysis involves  analyzing the performance under node churn~\cite{pandurangan2003building, augustine2015enabling}, with limited peer addresses known at each node (that are dynamically updated as part of a peer-discovery protocol).
